\documentstyle[amssymb,preprint,aps]{revtex}

\tightenlines

\begin{document}
\title{Quantum Bit Commitment Using Entangled States}
\author{Guang-Ping He}
\address{Advanced Research Center, Zhongshan University, Guangzhou 510275, China}
\maketitle

\begin{abstract}
Based on the fact that the entanglement can not be created locally, we
proposed a quantum bit commitment protocol, in which entangled states and
quantum algorithms is used. The bit is not encoded with the form of the
quantum states, and delaying the measurement is required. Therefore the
protocol will not be denied by the Mayers-Lo-Chau no-go theorem, and
unconditional security is achieved.
\end{abstract}

\pacs{03.67.-a, 89.70.+c, 03.65.-w}

Started from the original idea of Wiesner \cite{Wiesner}, quantum
cryptography is playing an essential role in nowadays research on quantum
information. Besides the well-known quantum key distribution protocol \cite
{BB84,Ekert91,B92}, another crucial primitive in quantum cryptography is
quantum bit commitment (QBC). As shown by Yao \cite{Yao}, a secure QBC
scheme can be used to implement a secure quantum oblivious transfer scheme 
\cite{BBCS92,OT}. And Kilian \cite{Kilian} has shown that, in classical
cryptography, oblivious transfer can be used to implement two-party secure
computations \cite{Crepeau95}. Many other cryptographies, such as quantum
coin tossing \cite{BB84,CT1,CT2,CT3} and quantum oblivious mutual
identification \cite{OMI}, can also be constructed over QBC. All these
cryptographies are very useful in the so-called ``post-cold-war era'', with
a wide range of economic, financial and even military applications. In
classical cryptography, these tasks can only be done through trusted
intermediaries. Otherwise some unproven computational assumptions must be
invoked, such as the hardness of factoring, which can easily be broken when
quantum computer becomes practical \cite{Shor}. Therefore people hope that
quantum cryptography can get rid of those requirements, and the same goals
can be achieved using the laws of physics alone. However, Mayers, Lo and
Chau have claimed that unconditionally secure QBC scheme can never be
achieved in principle \cite{Mayers,Lo}, and all the protocols formerly
purposed \cite{Yao,BCJL93} are insecure. By delaying the measurement on
quantum states, the participants can always succeed in cheating with
Einstein-Podolsky-Rosen(EPR)-type of attacks, or the so-called Mayers
attacks. This discovery is called the Mayers-Lo-Chau no-go theorem or MLC
theorem. During the past half decade, attempts on fixing the problem with
classical BC protocols (such as the classical computational BC scheme \cite
{Naor90,NOVY92} or the two-prover BC scheme \cite{twoprover}) are also
proven to be failed later \cite{LANL,PhysicaD}. Some protocols have to rely
on relativity assumptions \cite{relativity} or reduce to conditionally
secure \cite{conditional}. The failure of QBC seems to bring a fatal
limitation to the power of quantum cryptography.

But in this paper, we will propose a new QBC protocol which can not be
denied by the MLC theorem. EPR attacks will no longer succeed and
unconditional security can be achieved.

A bit commitment scheme between two parties (Alice and Bob) generally
includes two phases. In the commit phase, Alice has in mind a bit ($b=0$ or $%
1$) which she wants to commit to Bob. So she sends him a piece of evidence.
Later, in the unveil phase, Alice announces the value of $b$, and Bob checks
it with the evidence. A protocol is said to be binding if Alice cannot
change the value of $b$ after the commit phase, and is said to be concealing
if Bob cannot tell what $b$ is before the unveil phase. A secure protocol
needs to be both binding and concealing.

The argument of the MLC theorem is based on the Yao's general model of QBC 
\cite{Yao}. According to this model, previously proposed protocols that
proven to be insecure are all starting with the following steps: Alice
prepares a state $\left| 0\right\rangle =\sum\limits_{j}\sqrt{\lambda _{j}}%
\left| \alpha _{j}\right\rangle \otimes \left| \beta _{j}\right\rangle $ if $%
b=0$ or $\left| 1\right\rangle =\sum\limits_{j}\sqrt{\lambda _{j}}\left|
\alpha _{j}^{\prime }\right\rangle \otimes \left| \beta _{j}\right\rangle $
if $b=1$, and sends the second register to Bob. Then Alice is supposed to
carry out measurement on the first register to make it collapse to $\left|
\alpha _{k}\right\rangle $ or $\left| \alpha _{k}^{\prime }\right\rangle $
according to the value of $b$. And Bob measures the second register to
verify Alice's commitment. But in these protocols, the entanglement inside
the quantum states is not fully utilized. Any classical information that the
participants need to announce during the commit phase required by the
protocol can all be calculated without the help of the entanglement. That
is, the calculation involved is not thoroughly a quantum algorithm. The
value of $b$ in fact depends only on the form of the first register, and not
the form of the entire entangled state. As we know, entangled quantum states
have the power to carry out parallel computations, which is much more
powerful than classical algorithm. Therefore it is not surprise to see that
by making full use of the entangled states, Alice can execute the protocol
successfully even she delays her measurement, just as if she is executing
the commitment with $b=0$ and $b=1$ simultaneously. Then in the unveil
phase, she can apply local transformation on the first register to map the
state between $\left| 0\right\rangle $ and $\left| 1\right\rangle $. This is
the reason why Alice can cheat in these protocols \cite{Mayers,Lo,Mayers2}.

So we can see that, to propose a secure QBC protocol that can stand this
so-called Mayers attack, we must make full use of the computation power of
the entangled states. The effect of the entanglement must be taken into
consideration throughout the commitment, thus quantum algorithms must be
involved. So the outline of our new protocol goes as: In the commit phase,
Alice and Bob first share some certain entangled states which can solve a
certain problem with quantum algorithm; Then Alice shows Bob that she has
indeed solved the problem. Solving the problem should be able to force Alice
to measure a minimum set of states even with the most efficient quantum
algorithm, while the other states can be left unmeasured. Then we correlate
the commit bit $b$ with the states according to whether the states is
measured or not. In the unveil phase, Alice should show Bob that there is a
certain number of states which are indeed unmeasured. A state which is
already measured and collapsed can not be used to fake a state which is
still entangled with another state. Therefore the security of the commitment
can be guaranteed.

For concreteness, in the following we shall use four quantum states of
photon with different polarizations in the description. But in fact the
protocol can be constructed on any other type of nonorthogonal states as
well. Here we denote the four states of light polarization of angles $%
0^{\circ }$, $45^{\circ }$, $90^{\circ }$\ and $135^{\circ }$ as $\left|
0,0\right\rangle $, $\left| 1,0\right\rangle $, $\left| 0,1\right\rangle $
and $\left| 1,1\right\rangle $\ respectively. We will also consider the
ideal setting only, where the quantum communication channel is supposed to
be error-free. Before we get to the protocol, let us first consider the
following problem:

\bigskip

{\it Problem P}{\sl :}

Alice and Bob execute the following procedure:

(1) Alice sends Bob a series of photons $\{\beta _{i}|i\in S\}$ where $%
S\equiv \{1,...,s\}$ is a set of natural numbers;

(2) $\stackrel{s}{%
\mathrel{\mathop{DO}\limits_{i=1}}%
}$\ Bob randomly picks a bit $p_{i}^{\prime }$ and measures $\beta _{i}$ in
the rectilinear basis ($0^{\circ }$\ and $90^{\circ }$ polarized) if $%
p_{i}^{\prime }=0$, or the diagonal basis ($45^{\circ }$\ and $135^{\circ }$
polarized) if $p_{i}^{\prime }=1$. The outcome is denoted as $\left|
p_{i}^{\prime },q_{i}^{\prime }\right\rangle _{\beta }$;

(3) Bob announces to Alice a series of ``fake'' results $\{\left|
p_{i}^{\prime \prime },q_{i}^{\prime \prime }\right\rangle _{\beta }|i\in
S\} $, which needs not to be agreed with $\{\left| p_{i}^{\prime
},q_{i}^{\prime }\right\rangle _{\beta }|i\in S\}$. He can choose to apply
three types of lies: 
\begin{eqnarray*}
lie\text{ }a &:&\text{ }p_{i}^{\prime \prime }=p_{i}^{\prime }\wedge
q_{i}^{\prime \prime }=\urcorner q_{i}^{\prime }; \\
lie\text{ }b &:&\text{ }p_{i}^{\prime \prime }=\urcorner p_{i}^{\prime
}\wedge q_{i}^{\prime \prime }=q_{i}^{\prime }; \\
lie\text{ }c &:&\text{ }p_{i}^{\prime \prime }=\urcorner p_{i}^{\prime
}\wedge q_{i}^{\prime \prime }=\urcorner q_{i}^{\prime }.
\end{eqnarray*}
Let $L_{a}=\{i\in S|$ $\left| p_{i}^{\prime \prime },q_{i}^{\prime \prime
}\right\rangle _{\beta }=$ $\left| p_{i}^{\prime },\urcorner q_{i}^{\prime
}\right\rangle _{\beta }\}$, $L_{b}=\{i\in S|$ $\left| p_{i}^{\prime \prime
},q_{i}^{\prime \prime }\right\rangle _{\beta }=$ $\left| \urcorner
p_{i}^{\prime },q_{i}^{\prime }\right\rangle _{\beta }\}$, and $L_{c}=\{i\in
S|$ $\left| p_{i}^{\prime \prime },q_{i}^{\prime \prime }\right\rangle
_{\beta }=$ $\left| \urcorner p_{i}^{\prime },\urcorner q_{i}^{\prime
}\right\rangle _{\beta }\}$, with $f_{a}=\left| L_{a}\right| /s$, $%
f_{b}=\left| L_{b}\right| /s$ and $f_{c}=\left| L_{c}\right| /s$ denoting
the frequencies of Bob applying each type of lies. Suppose that $0<f_{a},$ $%
f_{b},$ $f_{c}<1/4$ and $f_{b}>f_{c}$.

Now the question is: how can Alice detects a set of lies $D$ such that $%
D\subset L_{a}\cup L_{b}\cup L_{c}$ and $\left| D\right| \sim
(f_{a}/2+f_{b}/4+f_{c}/4)s$?

\bigskip

This problem can easily be solved by the following ``semi-classical''
method. Alice can determine the states of all the $s$ photons beforehand,
i.e. in step (1) she prepares every photon $\beta _{i}$ in a pure state $%
\left| p_{i},q_{i}\right\rangle _{\beta }$ non-entangled with any other
systems. Then she sets $D=\{i\in S|\left| p_{i}^{\prime \prime
},q_{i}^{\prime \prime }\right\rangle _{\beta }=\left| p_{i},\urcorner
q_{i}\right\rangle _{\beta }\}$ after Bob announces $\{\left| p_{i}^{\prime
\prime },q_{i}^{\prime \prime }\right\rangle _{\beta }|i\in S\}$\ in step
(3). Now let us evaluate the size of this set $D$. Since $p_{i}^{\prime }$
is randomly chosen by Bob, for half of the states Bob will by chance choose
the correct basis $p_{i}^{\prime }=p_{i}$. Among this half, whenever Bob
applies a $lie$ $a$, it will be detected by Alice since she knows that Bob
can never find $q_{i}$ as $\urcorner q_{i}$ in his measurement once he uses
the correct basis. But no $lie$ $b$ and $lie$ $c$ will be detected, since
when Bob announces $p_{i}^{\prime \prime }=\urcorner p_{i}^{\prime
}=\urcorner p_{i}$, Alice does not know what the result should be when $%
\left| p_{i},q_{i}\right\rangle _{\beta }$\ is measured in the wrong basis $%
p_{i}^{\prime \prime }$. Meanwhile for the other half of states that Bob has
measured with the wrong basis $p_{i}^{\prime }=\urcorner p_{i}$, the
probabilities of finding $q_{i}^{\prime }=q_{i}$\ and $q_{i}^{\prime
}=\urcorner q_{i}$\ are both $1/2$. Therefore when Bob applies $lie$ $b$ or $%
lie$ $c$, $p_{i}^{\prime \prime }=\urcorner p_{i}^{\prime }$ becomes the
correct basis. The probability for such a state to satisfy $\left|
p_{i}^{\prime \prime },q_{i}^{\prime \prime }\right\rangle _{\beta }=\left|
p_{i},\urcorner q_{i}\right\rangle _{\beta }$\ is then $1/2$. But no $lie$ $%
a $ will be detected in this case since $p_{i}^{\prime \prime
}=p_{i}^{\prime } $ is the wrong basis now. So we can see that, the number
of lies that Alice totally detects is $\left| D\right| \sim
(f_{a}/2+f_{b}/4+f_{c}/4)s$. That is, such a set $D$ is just what is
required by the problem.

But with full quantum algorithms we can solve the problem more efficiently.
Alice can prepare every photon $\beta _{i}$ as a mixture entangled with
another system $\alpha _{i}$. For example, in step (1) she can prepare the
state of the whole incremental system as $\left| \psi _{i}\right\rangle
=\left| \alpha _{i}\otimes \beta _{i}\right\rangle =\cos \theta _{i}\left|
x\right\rangle _{\alpha }\otimes \left| 0,q_{i}\right\rangle _{\beta }+\sin
\theta _{i}\left| y\right\rangle _{\alpha }\otimes \left|
1,q_{i}\right\rangle _{\beta }$, where $\left| x\right\rangle _{\alpha }$
and $\left| y\right\rangle _{\alpha }$ are orthogonal to each other, $%
q_{i}\in \{0,1\}$ and $\theta _{i}\in (0,\pi /2)$. She sends $\beta _{i}$ to
Bob, and after step (3) she divides $S$ \ into two subsets: $M=\{i\in
S|q_{i}^{\prime \prime }=\urcorner q_{i}\}$ and $U=\{i\in S|q_{i}^{\prime
\prime }=q_{i}\}$. Due to the specific form of $\left| \psi
_{i}\right\rangle $, Bob is more likely to find $\beta _{i}$ as $q_{i}$ than 
$\urcorner q_{i}$\ no matter which basis he uses. Therefore when Alice finds
Bob announcing $q_{i}^{\prime \prime }=\urcorner q_{i}$, she knows that he
is more likely to be lying. So for the states whose indices are included in $%
U$, she can just leave them unmeasured. And for $\forall i\in M$, She
measures $\alpha _{i}$ in the basis $(\left| x\right\rangle _{\alpha
},\left| y\right\rangle _{\alpha })$. She sets $p_{i}=0$ if she finds $%
\left| x\right\rangle _{\alpha }$ or $p_{i}=1$ if she finds $\left|
y\right\rangle _{\alpha }$. Then she sets $D=\{i\in M|p_{i}=p_{i}^{\prime
\prime }\}$. Detailed analyses can prove that both $D\subset L_{a}\cup
L_{b}\cup L_{c}$ and $\left| D\right| \sim (f_{a}/2+f_{b}/4+f_{c}/4)s$ are
automatically satisfied. Calculations also show that $\left| M\right| \sim
\lbrack 1/4+(f_{a}+f_{c})/2]s$ when Bob chooses $p_{i}^{\prime }$ randomly.
Thus we see that Alice can detect $D$ by measuring only $%
[1/4+(f_{a}+f_{c})/2]s$\ states. In the ``semi-classical'' method described
above, Alice's action in step (1) is equivalent to preparing the states in
an entangled form as well at first , but then measures all the $s$ entangled
states $\left| \psi _{i}\right\rangle $ to make $\beta _{i}$ collapse into
non-entangled pure states before Bob measure them. So we can see now with
the full use of the computational power of the entangled states, Alice
manages to measure less states than the ``semi-classical'' method while the
same goal is achieved.

This quantum algorithm is already the most efficient one. One can verify
that preparing $\left| \psi _{i}\right\rangle $ in other forms will have to
measure more states when detecting $D$. For example, if Alice prepares $%
\left| \psi _{i}\right\rangle =\left| \alpha _{i}\otimes \beta
_{i}\right\rangle =\cos \theta _{i}\left| x\right\rangle _{\alpha }\otimes
\left| 0,0\right\rangle _{\beta }+\sin \theta _{i}\left| y\right\rangle
_{\alpha }\otimes \left| 0,1\right\rangle _{\beta }$ and always measures $%
\alpha _{i}$ in the basis which can force $\beta _{i}$ to collapse to $%
p_{i}^{\prime \prime }$, she will need to measure $s/2$ states to detect $D$%
. Or if she prepares $\left| \psi _{i}\right\rangle =\left| \alpha
_{i}\otimes \beta _{i}\right\rangle =\cos \theta _{i}\left| x\right\rangle
_{\alpha }\otimes \left| 0,0\right\rangle _{\beta }+\sin \theta _{i}\left|
y\right\rangle _{\alpha }\otimes \left| 1,1\right\rangle _{\beta }$ and
always measures those that satisfy $\left| p_{i}^{\prime \prime
},q_{i}^{\prime \prime }\right\rangle _{\beta }=\left| 0,1\right\rangle
_{\beta }\vee \left| p_{i}^{\prime \prime },q_{i}^{\prime \prime
}\right\rangle _{\beta }=\left| 1,0\right\rangle _{\beta }$, she will need
to measure $[1/4+(f_{a}+f_{b})/2]s$ states. All these numbers are larger
than $[1/4+(f_{a}+f_{c})/2]s$ given $f_{a},$ $f_{b},$ $f_{c}<1/4$ and $%
f_{b}>f_{c}$.

So if we build a protocol in which Alice is required to solve {\it Problem P}
while only $[1/4+(f_{a}+f_{c})/2]s$ states are allowed to be measured, she
has to follow the above quantum algorithm honestly. Now let us give a
parameter $c_{i}^{0}$ to each state $\left| \psi _{i}\right\rangle $ ($i\in
S-D$), and set $c_{i}^{0}=0$ if $i\in U$ which means $\alpha _{i}$ is
unmeasured by Alice, or $c_{i}^{0}=1$ if $i\in M-D$ which means $\alpha _{i}$
is already measured by Alice. Thus after solving {\it Problem P}, a string $%
c^{0}=(c_{1}^{0}c_{2}^{0}...c_{n}^{0})$ ($n\equiv \left| S-D\right| $) is
created. Then we can adopt the codeword method in BCJL QBC protocol \cite
{BCJL93}, encoding a codeword with $c^{0}$ to make it oriented to the commit
bit $b$. So the entire description of our QBC protocol is:

\bigskip

The commit protocol: ($commit(b)$)

(C1) Alice and Bob first agree on a security parameter $s$, then $\stackrel{s%
}{%
\mathrel{\mathop{DO}\limits_{i=1}}%
}$ Alice picks $\theta _{i}\in (0,\pi /2)$ ($\theta _{i}$ needs not to be
different for different $i$. For example, one can fix $\theta _{i}=\pi /4$\
throughout the whole protocol) and randomly picks $q_{i}\in \{0,1\}$, and
prepares an entangled state $\left| \psi _{i}\right\rangle =\left| \alpha
_{i}\otimes \beta _{i}\right\rangle =\cos \theta _{i}\left| x\right\rangle
_{\alpha }\otimes \left| 0,q_{i}\right\rangle _{\beta }+\sin \theta
_{i}\left| y\right\rangle _{\alpha }\otimes \left| 1,q_{i}\right\rangle
_{\beta }$. Then she sends $\beta _{i}$ to Bob and stores $\alpha _{i}$;

(C2) Bob chooses a number $s^{\prime }$ ($0\leqslant s^{\prime }\leqslant s$%
) and randomly divides $S\equiv \{1,...,s\}$ into two subsets $S^{\prime }$
and $S^{\prime \prime }$ such that $\left| S^{\prime }\right| =s^{\prime }$, 
$S^{\prime \prime }=S-S^{\prime }$. Then for $\forall i\in S^{\prime }$ Bob
stores $\beta _{i}$\ unmeasured. And for $\forall i\in S^{\prime \prime }$
Bob randomly picks a basis $p_{i}^{\prime }$ and measures $\beta _{i}$. The
outcome is denoted as $\left| p_{i}^{\prime },q_{i}^{\prime }\right\rangle
_{\beta }$;

(C3) Bob chooses $f_{a}$, $f_{b}$, $f_{c}$ ($0<f_{a},$ $f_{b},$ $f_{c}<1/4$
and $f_{b}>f_{c}$) and announces to Alice the ``fake'' results $\{\left|
p_{i}^{\prime \prime },q_{i}^{\prime \prime }\right\rangle _{\beta }|i\in
S\} $ such that $f_{a}=(\left| L_{a}\right| +s^{\prime }/4)/s$, $%
f_{b}=(\left| L_{b}\right| +s^{\prime }/4)/s$ and $f_{c}=(\left|
L_{c}\right| +s^{\prime }/4)/s$, where $L_{a}=\{i\in S^{\prime \prime }|$ $%
\left| p_{i}^{\prime \prime },q_{i}^{\prime \prime }\right\rangle _{\beta }=$
$\left| p_{i}^{\prime },\urcorner q_{i}^{\prime }\right\rangle _{\beta }\}$, 
$L_{b}=\{i\in S^{\prime \prime }|$ $\left| p_{i}^{\prime \prime
},q_{i}^{\prime \prime }\right\rangle _{\beta }=$ $\left| \urcorner
p_{i}^{\prime },q_{i}^{\prime }\right\rangle _{\beta }\}$, and $L_{c}=\{i\in
S^{\prime \prime }|$ $\left| p_{i}^{\prime \prime },q_{i}^{\prime \prime
}\right\rangle _{\beta }=$ $\left| \urcorner p_{i}^{\prime },\urcorner
q_{i}^{\prime }\right\rangle _{\beta }\}$;

(C4) Alice divides $S$ \ into two subsets: $M=\{i\in S|q_{i}^{\prime \prime
}=\urcorner q_{i}\}$ and $U=\{i\in S|q_{i}^{\prime \prime }=q_{i}\}$. For $%
\forall i\in M$, She measures $\alpha _{i}$ in the basis $(\left|
x\right\rangle _{\alpha },\left| y\right\rangle _{\alpha })$. She sets $%
p_{i}=0$ if she finds $\left| x\right\rangle _{\alpha }$ or $p_{i}=1$ if she
finds $\left| y\right\rangle _{\alpha }$. Then she sets $D=\{i\in
M|p_{i}=p_{i}^{\prime \prime }\}$ announces it to Bob;

(C5) Bob sets $D_{s^{\prime }}=D\cap S^{\prime }$. Then he measures $\beta
_{i}$\ ($\forall i\in D_{s^{\prime }}$) in the basis $p_{i}^{\prime
}=p_{i}^{\prime \prime }$ and denotes the outcome as $\left| p_{i}^{\prime
},q_{i}^{\prime }\right\rangle _{\beta }$. He agrees to continue only if $%
\{i\in D_{s^{\prime }}|\left| p_{i}^{\prime },q_{i}^{\prime }\right\rangle
_{\beta }=\left| p_{i}^{\prime \prime },q_{i}^{\prime \prime }\right\rangle
_{\beta }\}=\phi $, $D\subset L_{a}\cup L_{b}\cup L_{c}\cup S^{\prime }$ and 
$\left| D\right| \sim (f_{a}/2+f_{b}/4+f_{c}/4)s$;

(C6) Alice sets $c_{i}^{0}=0$ if $i\in U$ or $c_{i}^{0}=1$ if $i\in M-D$.
Thus she obtains a binary string $c^{0}=(c_{1}^{0}c_{2}^{0}...c_{n}^{0})$ ($%
n\equiv \left| S-D\right| $);

(C7) Alice and Bob execute the BCJL protocol \cite{BCJL93} by using $c^{0}$
to encode the codeword ($c^{0}$ itself is not announced to Bob). That is:

\qquad (C7.1) Bob chooses a Boolean matrix $G$ as the generating matrix of a
binary linear $(n,k,d)$-code $C$ and announces it to Alice, where the ratios 
$d/n$ and $k/n$ are agreed on by both Alice and Bob;

\qquad (C7.2) Alice chooses a non-zero random $n$-bit string $%
r=(r_{1}r_{2}...r_{n})\in \{0,1\}^{n}$ and announces it to Bob;

\qquad (C7.3) Now Alice has in mind the value of the bit $b$ that she wants
to commit. Then she chooses a random $n$-bit codeword $%
c=(c_{1}c_{2}...c_{n}) $ from $C$ such that $c\odot r=b$ (Here $c\odot
r\equiv \bigoplus\limits_{i=1}^{n}c_{i}\wedge r_{i}$);

\qquad (C7.4) Alice announces to Bob $c^{\prime }=c\oplus c^{0}$.

\bigskip

The unveil protocol: ($unveil(b,c,c^{0},\left| \psi _{i}\right\rangle )$)

(U1) Alice announces $b$, $c$, $c^{0}$, $\{q_{i},\theta _{i}|$ $i\in S\}$
and $\{p_{i}|i\in M\}$\ to Bob;

(U2) Alice sends the quantum registers $\{\alpha _{i}|i\in U\}$ to Bob;

(U3) Bob finishes the measurement on $\{\alpha _{i}|i\in U\}$ and $\{\beta
_{i}|i\in S^{\prime }\}$ to check Alice's announcement;

(U4) Bob checks $\left| M\right| \sim \lbrack 1/4+(f_{a}+f_{c})/2]s$ and $%
(M-D)\cap L_{b}=\phi $;

(U5) Bob checks $b=c\odot r$ and ($c$ is a codeword).

\bigskip

Unlike those described in {\it Problem P}, in step (C2) we allow Bob to
choose a subset $S^{\prime }$, and delay the measurement on $\beta _{i}$\ ($%
\forall i\in S^{\prime }$). For the states in this set, since Bob has to
announce $\left| p_{i}^{\prime \prime },q_{i}^{\prime \prime }\right\rangle
_{\beta }$\ randomly before he obtains $\left| p_{i}^{\prime },q_{i}^{\prime
}\right\rangle _{\beta }$, it is equivalent to lie with the frequencies $%
f_{a}=f_{b}=f_{c}=1/4$ among the set. Thus by choosing $s^{\prime }$
properly, Bob can still control the total lying frequencies $f_{a}$, $f_{b}$
and $f_{c}$, as those described in step (C3). The purpose of $S^{\prime }$
is to enhance Bob's chance on catching Alice cheating in steps (C5) and
(U3). However, the protocol is still valid even if Bob chooses $S^{\prime
}=\phi $.

The purpose of step (U3) is to make sure that Alice does not shift the bits
in string $c^{0}$ from $1$ to $0$. In another word, it is to check whether
Alice has already measured a state $\alpha _{i}$ to make $\beta _{i}$
collapse, but still tries to say that the two states are left entangled.
There are many type of measurement that Bob can perform to catch this kind
of cheating. When both the two registers $\alpha _{i}$\ and $\beta _{i}$ in $%
\left| \psi _{i}\right\rangle $\ are not measured before (i.e. $i\in U\cap
S^{\prime }$), Bob can simply sort them by $\theta _{i}$ and $q_{i}$\ and
then measure the amount of entanglement \cite{entanglement} between them.
Since local transformations will not affect the entanglement, Alice can not
make a measured $\alpha _{i}$\ entangle with $\beta _{i}$ without the help
from Bob. So if the result of Bob's measurement turns out to be zero or much
different from the expected value calculated from the form of $\left| \psi
_{i}\right\rangle $\ Alice announced, Bob should reject this commitment.

For the other states where one of the registers of $\left| \psi
_{i}\right\rangle $\ is already measured in the commit phase, Bob can use
the form of $\left| \psi _{i}\right\rangle $\ Alice announced to calculate
the expected state $\left| e_{i}\right\rangle $\ to which the other register
of $\left| \psi _{i}\right\rangle $\ should\ collapse. Then he measures this
register in the basis $(\left| e_{i}\right\rangle ,\left| e_{i}\right\rangle
^{\perp })$. As we know, different measured results of one of the registers
will cause the other register to collapse to different states, and these
states are not orthogonal to each other when $\theta _{i}\neq 0\wedge \theta
_{i}\neq \pi /2$. Therefore if Alice has not followed the protocol honestly,
the unmeasured register will have a non-zero probability to be found as $%
\left| e_{i}\right\rangle ^{\perp }$ by Bob. For instance, suppose Alice has
formerly prepared a state as $\left| \psi _{i_{0}}\right\rangle =\left|
\alpha _{i_{0}}\otimes \beta _{i_{0}}\right\rangle =1/\sqrt{2}(\left|
x\right\rangle _{\alpha }\otimes \left| 0,0\right\rangle _{\beta }+\left|
y\right\rangle _{\alpha }\otimes \left| 1,0\right\rangle _{\beta })$. And in
step (C3) Bob announces $\left| p_{i_{0}}^{\prime \prime },q_{i_{0}}^{\prime
\prime }\right\rangle _{\beta }=\left| 1,1\right\rangle _{\beta }$. Alice
will then include the index $i_{0}$ of this state in set $M$ and measures $%
\alpha _{i_{0}}$ in the basis $(\left| x\right\rangle _{\alpha },\left|
y\right\rangle _{\alpha })$. Suppose that she obtain $\left| x\right\rangle
_{\alpha }$\ in her measurement. So she will not include $i_{0}$ in set $D$.
Now since $i_{0}\in M-D$, she should set $c_{i_{0}}^{0}=1$. However, the
dishonest Alice wants Bob to believe $c_{i_{0}}^{0}=0$, so she must send Bob
a fake state $\tilde{\alpha}_{i_{0}}$. But she does not know the result of
Bob's measurement on $\beta _{i_{0}}$. Since she has found $\alpha _{i_{0}}$
as $\left| x\right\rangle _{\alpha }$\ in her measurement, there are three
possibilities: Bob has found $\beta _{i_{0}}$ as $\left| 0,0\right\rangle
_{\beta }$, $\left| 1,0\right\rangle _{\beta }$, or $\left| 1,1\right\rangle
_{\beta }$. Then $\alpha _{i_{0}}$ has collapsed to $\sqrt{2/3}\left|
x\right\rangle _{\alpha }+\sqrt{1/3}\left| y\right\rangle _{\alpha }$, $%
\sqrt{1/3}\left| x\right\rangle _{\alpha }+\sqrt{2/3}\left| y\right\rangle
_{\alpha }$, or $\left| x\right\rangle _{\alpha }$ respectively. If she
prepares $\left| \tilde{\alpha}_{i_{0}}\right\rangle =\left| x\right\rangle
_{\alpha }$ and sends to Bob, chances are that Bob has formerly obtained $%
\left| p_{i_{0}}^{\prime },q_{i_{0}}^{\prime }\right\rangle _{\beta }=\left|
0,0\right\rangle _{\beta }$ in step (C2) so he is expecting $\left| \alpha
_{i_{0}}\right\rangle =\sqrt{2/3}\left| x\right\rangle _{\alpha }+\sqrt{1/3}%
\left| y\right\rangle _{\alpha }$. Then when he measures $\tilde{\alpha}%
_{i_{0}}$ in the basis $(\sqrt{2/3}\left| x\right\rangle _{\alpha }+\sqrt{1/3%
}\left| y\right\rangle _{\alpha },-\sqrt{1/3}\left| x\right\rangle _{\alpha
}+\sqrt{2/3}\left| y\right\rangle _{\alpha })$, he stands $1/3$ chances to
finds $\tilde{\alpha}_{i_{0}}$ as $-\sqrt{1/3}\left| x\right\rangle _{\alpha
}+\sqrt{2/3}\left| y\right\rangle _{\alpha }$ and catches Alice cheating. In
this case, the probability for Alice to cheat successfully for this single
bit is $f=2/3$. As the minimum distance between codewords is $d$, to keep
the total number of $0$ in $c^{0}$\ unchanged, a dishonest Alice will have
to shift at least $d/2$ bits of $c^{0}$ from $1$ to $0$ to fulfill her
cheating. Therefore the total probability for Alice to successfully cheat
this way without being caught is less than $\max (f)^{d/2}$. Since $d/n$ is
fixed to be a constant in the protocol and $n\varpropto s$, this probability
drops exponentially to zero as the security parameter $s$ increases.

The purpose of step (U4) is to make sure that Alice does not shift the bits
in string $c^{0}$ from $0$ to $1$. In our protocol, although Alice can shift
a bit $c_{i}^{0}$ from $0$ into $1$ simply by measuring $\alpha _{i}$, the
total number of $1$ in $c^{0}$\ is already restrained to be about $\left|
M-D\right| \sim (1-f_{b}+f_{c})s/4$. Since $\left| M\right| $ is already the
minimum of the number of states that Alice has to measure to solve {\it %
Problem P}, if she shift more bits from $0$ into $1$, there will be too much 
$1$\ in $c^{0}$. So this kind of cheating is easy for Bob to find out. Also,
solving {\it Problem P}\ with $\left| \psi _{i}\right\rangle =\left| \alpha
_{i}\otimes \beta _{i}\right\rangle =\cos \theta _{i}\left| x\right\rangle
_{\alpha }\otimes \left| 0,q_{i}\right\rangle _{\beta }+\sin \theta
_{i}\left| y\right\rangle _{\alpha }\otimes \left| 1,q_{i}\right\rangle
_{\beta }$ has a characteristic property: all $lie$ $b$ among the set of
states that Alice measured will be detected. This is because when $\beta
_{i} $ is found as $\left| 0,\urcorner q_{i}\right\rangle _{\beta }$\ (or $%
\left| 1,\urcorner q_{i}\right\rangle _{\beta }$) in Bob's measurement, $%
\alpha _{i} $\ will collapse to $\left| y\right\rangle _{\alpha }$\ (or $%
\left| x\right\rangle _{\alpha }$ respectively). If Bob applies $lie$ $b$ by
announcing it as $\left| 1,\urcorner q_{i}\right\rangle _{\beta }$\ (or $%
\left| 0,\urcorner q_{i}\right\rangle _{\beta }$), Alice is then expecting
to find $\alpha _{i}$ as $\left| x\right\rangle _{\alpha }$ (or $\left|
y\right\rangle _{\alpha }$)\ in her measurement. Since $\left|
x\right\rangle _{\alpha }$ and $\left| y\right\rangle _{\alpha }$\ are
orthogonal to each other, so when Alice measures only the states that
satisfy $q_{i}^{\prime \prime }=\urcorner q_{i}$, $lie$ $b$ will be 100\%
detected and none of them will be left in set $M-D$. Thus if Bob finds $%
(M-D)\cap L_{b}\neq \phi $\ in step (U4), he knows that Alice must have
measured some states which do not satisfy $q_{i}^{\prime \prime }=\urcorner
q_{i}$, or even has not prepared $\left| \psi _{i}\right\rangle $\ in the
correct form.

Therefore if Alice alters much of the bits in $c^{0}$, she will inevitably
be caught. Nevertheless, due to the fluctuation of random distribution, we
can not expect the size of $D$ detected by Alice to be exactly equal to $%
(f_{a}/2+f_{b}/4+f_{c}/4)s$. So if Alice alters only few bits of $c^{0}$,
she may escape from being caught. But the codeword method in the BCJL
protocol can avoid this situation. That is, since the minimum distance
between any legal codewords is $d$, altering only a small number of bits of $%
c^{0}$ will not be enough to change a codeword into another legal codeword.
Therefore this way of cheating will make no sense to Alice at all.

Now we will show that the protocol is also secure against Bob. During the
commit phase, since $q_{i}$\ is kept secret by Alice, Bob can not know how
to divide $S$ into subsets $M$ and $U$. Though he knows that in the $n$-bit
string $c^{0}$ ($n=\left| S-D\right| \sim (1-f_{a}/2-f_{b}/4-f_{c}/4)s$),\
there are $d^{0}\equiv \left| M-D\right| \sim (1-f_{b}+f_{c})s/4$ bits in $%
c^{0}$ take the value $1$, and the other $(n-d^{0})$\ bits are $0$, he does
not know the position of these bits. Thus the possible number of $c^{0}$ is $%
{n \choose d^{0}}%
$. Then Theorem 3.4 in Ref.\cite{BCJL93} applies. Briefly, as $d^{0}>\gamma
n $ ($\gamma \equiv H^{-1}(1/2)\sim 0.1100279$), we have $%
{n \choose d^{0}}%
>%
{n \choose \gamma n}%
\geqslant 2^{n/2}/\sqrt{n}$. Divide by $2^{n-k}$ (the number of syndromes of
the code $C$), and we get: the number of codewords at Hamming distance $%
d^{0} $ has a lower bound $2^{k-n/2}/\sqrt{n}$, which is exponentially large
in $n$ as long as we choose $k/n>1/2$ in step (C7.1). Therefore Lemmas 3.5
and 3.6 of Ref.\cite{BCJL93} are also valid for our protocol. That is, Bob
has exponentially small amount of Shannon information on the value of $b$
before the unveil phase.

So we can see that our protocol is both unconditionally binding and
concealing, therefore it is unconditionally secure. Briefly, the protocol
evades the MLC-theorem for the following reason. There are two tasks for
Alice to accomplish during the commit phase: {\it Task 1}: solve {\it %
Problem P}; and {\it Task 2}: commit the bit $b$. The purpose of {\it Task 1}
is to prepare the input states for {\it Task 2}. The form of {\it Task 2} is
quite similar to the BCJL QBC protocol. However, there is a critical
difference: the encoding method. Unlike any protocols that can concluded by
the Yao's general QBC model, in our protocol, whether a state $\left| \psi
_{i}\right\rangle =\left| \alpha _{i}\otimes \beta _{i}\right\rangle $ is
encoded as $0$ or $1$ is not depended on the form of $\alpha _{i}$, but on
whether $\left| \psi _{i}\right\rangle $\ is an entangled state or not. If $%
\left| \psi _{i}\right\rangle $ can be written as $\left| \alpha
_{i}\right\rangle \otimes \left| \beta _{i}\right\rangle $\ (which means
that it is a non-entangled product state) we take $c_{i}^{0}=1$, else we
take $c_{i}^{0}=0$. Since it is a basic principle that the entanglement
between two systems $\alpha _{i}$ and $\beta _{i}$\ can not be created
locally, there does not exist any local unitary transformation for Alice to
map a state $\left| \psi _{i}\right\rangle =\left| \alpha _{i}\right\rangle
\otimes \left| \beta _{i}\right\rangle $\ into an entangled state. Of course
if Alice can maintain every input state of {\it Task 2} in an entangled form 
$\left| \psi _{i}\right\rangle =\left| \alpha _{i}\otimes \beta
_{i}\right\rangle $, she can unveil $c_{i}^{0}$ with any value she like,
since such a state is free to map into $\left| \psi _{i}\right\rangle
=\left| \alpha _{i}\right\rangle \otimes \left| \beta _{i}\right\rangle $.
But to accomplish {\it Task 1}, Alice inevitably has to measure at least a
certain number of these states to break down the entanglement between $\beta
_{i}$ and any other systems and make $\left| \psi _{i}\right\rangle $
collapse to $\left| \alpha _{i}\right\rangle \otimes \left| \beta
_{i}\right\rangle $ (Here $\alpha _{i}$ can represent any systems not on
Bob's side, including the environment). And this number sets the maximum of
the allowed number of $1$ in the codeword string $c^{0}$ in our protocol.
Therefore, no LOCAL unitary transformation will be available for Alice to
map the state $\left| b\right\rangle =\left| \psi _{1}\right\rangle \otimes
\left| \psi _{2}\right\rangle \otimes ...\otimes \left| \psi
_{n}\right\rangle $ into $\left| \urcorner b\right\rangle $. By this means,
the cheating strategy in the MLC theorem can not work any more, and
unconditionally secure is achieved. Full mathematical proof and detailed
discussion on the limitation of the MLC theorem will be supplied elsewhere.

Thus by using entangled states to run quantum algorithms, we propose an
unconditionally secure quantum bit commitment protocol. Therefore all the
other cryptographies that base on bit commitment, such as unconditionally
secure quantum oblivious transfer, two-party secure computations, quantum
coin tossing and quantum oblivious mutual identification are then straight
forward. The potential of quantum cryptography meets a great development.

\end{document}